\documentclass[12pt,preprint]{aastex}
\usepackage{color}
\usepackage{ulem}

\shorttitle{kHz QPOs from XTE~J1701--462}
\shortauthors{Barret et al.}

\def\xtej{XTE~J1701--462}
\begin{document}

\title{The drop of the coherence of the lower kHz QPOs is also observed 
in XTE J1701-462}

\author{D. Barret\altaffilmark{1} and M. Bachetti\altaffilmark{1}}
\affil{Universit\'e de Toulouse (UPS), 9 av. du Colonel Roche, 31028 
Toulouse Cedex 9, France; Centre National de la Recherche Scientifique, 
Centre d'Etude Spatiale des Rayonnements, UMR 5187, 9 av. du Colonel 
Roche, BP 44346, 31028 Toulouse Cedex 4, France}

\author{M. Coleman Miller\altaffilmark{2}}
\affil{Department of Astronomy and Maryland Astronomy Center 
for Theory and Computation, University of Maryland, College Park, 
MD 20742-2421, USA.  Also Joint Space Science Institute.}
\email{didier.barret@cesr.fr}

\begin{abstract}
We investigate the quality factor and RMS amplitude of
the lower kHz QPOs from \xtej\, a unique X-ray source which
was observed in both the so-called Z and atoll states. Correcting
for the frequency drift of the QPO, we show that, as in all
sources for which such a correction can be applied, the quality
factor and RMS amplitude drops sharply above above a critical frequency. 
For \xtej\ this frequency is estimated to be $\sim 800$ Hz, where
the quality factor reaches a maximum of $\sim 200$ (e.g. a value
consistent with the one observed from more classical systems, such as
4U~1636--536). Such a drop has been interpreted as the signature of
the innermost stable circular orbit, and that interpretation is
consistent with the observations we report here.
The kHz QPOs in the Z state are much less coherent and lower
amplitude than they are in the atoll state. We argue that the
change of the QPO properties between the two source states is
related to the change of the scale height of the accretion disk; a
prediction of the toy model proposed by \citet{barret07mnras}.
As a by-product of our analysis, we also increased the significance
of the upper kHz QPO detected in the atoll phase up to $4.8\sigma$
(single trial significance), and show that the frequency
separation ($266.5\pm13.1$ Hz) is comparable with the one measured
from simultaneous twin QPOs the Z phase.

\end{abstract}



\section{Introduction}
Kilohertz quasi-periodic brightness variations (kHz QPOs) have been
detected from more than twenty low-mass X-ray binaries containing a
weakly magnetized neutron star \citep{vdk06}.  A pair of QPOs is
often detected, with a frequency separation that typically changes by
tens of Hertz as the individual QPO frequencies change by hundreds of
Hertz. In lower luminosity systems (often called atoll sources after
the tracks they make in color-color diagrams), the lower frequency
QPO can reach quality factors (Q$\equiv\nu/\Delta\nu$) up to 200,
whereas the upper kHz QPO is usually a broader feature with maximum Q
around 50 \citep[see,
e.g.][]{barret05amnras,barret05bmnras,barret06mnras}. In brighter
systems (often called Z sources), both the lower and upper kHz QPOs
are broader features, with maximum Q of a few tens
\citep{boutelier10mnras}. In lower luminosity sources, for which the
quality factor of the lower kHz QPO can be measured over its entire
frequency span after correcting for the frequency drifts, Q increases
with frequency until it reaches a maximum around 800--900 Hz, beyond
which a sharp drop-off is observed. The presence of the drop in many
systems, and its reproducibility in a given system  independent of
count rate and spectral hardness \citep{barret07mnras}, led to the
suggestion that it may be related to the existence of an innermost stable
circular orbit, a key prediction of strong field general relativity
\citep{barret06mnras}.

XTE~J1701--462 is a unique X-ray source, which
first behaved like a Z source at high luminosity, and
which later behaved like an atoll source at much lower luminosities.
The source was closely monitored with the RXTE Proportional Counter
Array during its 2006-2007 outburst
\citep{homan07apj,lin09apja,homan10apj}. Type I X-ray bursts were
studied by \citet{lin09apjb}. The three bursts observed occurred as
the source was in transition from the typical Z-source behavior to
the typical atoll-source behavior, at $\sim10$\% of the Eddington
luminosity. No significant burst oscillations in the range 30--4000 Hz
were found during these three bursts.   \citet{lin09apjb} also derive
a distance estimate of $8.8\pm1.3$ kpc from two radius expansion
bursts (the latter two of the three). In both states, kHz QPOs have
been reported, with very different properties, yet following the
general trend that in the Z state, QPOs showed lower Q and amplitude
than in the atoll state \citep{homan07apj,homan10apj,sanna10mnras}.
The luminosity range over which kHz QPOs are detected between the Z
and atoll state spans a factor of 15-20. The dramatic changes in the
QPO parameters between the two source states could not be due to
changes in the neutron star mass, its magnetic field, its spin, or
even the inclination of the accretion disk.  It is thus more 
plausible that they were
instead caused by a change in the properties of the accretion flow.
This led \citet{sanna10mnras} to conclude first that effects other
than the geometry of space time around the neutron star have a strong
influence on the coherence and amplitude of the kHz QPOs, and second
that the drop of the coherence and RMS amplitude of the lower kHz
QPOs, as we have observed it, could not be used to infer the
existence of the innermost stable circular orbit around a neutron
star.

In this paper, we revisit the RXTE observations reported by
\citet{sanna10mnras}, focussing on the lower kHz QPO, and apply the
same analysis procedures (frequency drift correction) as described
in \citet{barret06mnras}. The main reason is that, despite very few
detections overall (12 segments of observations in total), the
frequency span of the lower kHz QPO ranges from about 640 Hz to 850
Hz, and by comparison with other sources, this should be sufficient
to investigate the drop of its quality factor (no such drop is
obvious in Figure~3 of \citet{sanna10mnras} who used a different
analysis than  \citet{barret06mnras}). In the next section, we
describe the re-analysis of the lower kHz QPOs reported by
\citet{sanna10mnras} to show that the drop of its quality factor
and RMS amplitude is indeed observed. We then discuss
the changes in the properties of the QPOs from XTE~J1701--462 as
simply related to a likely decrease of the disk thickness between
the Z and atoll phases, as previously discussed in the framework of
the toy model presented in \citet{barret07mnras}.
\section{Observations}
We follow the definition of the Z and atoll states as in
\citet{sanna10mnras}: the separation is estimated around the end
of April 2007. We have retrieved from the HEASARC archive science 
event mode and
single bit data recorded by the RXTE Proportional Counter Array
(PCA). We consider data as segments of continuous observation (an ObsID 
may contain more than 1 segment). For each
segment, we have computed an average Power Density Spectrum (PDS)
with a 1 Hz resolution and an integration time of 16 seconds,
using events recorded between 2 and 40 keV. The PDS are normalized
according to \citet{Leahy:1983mb}, so that the Poisson noise level
is expected to be a constant close to 2. The PDS is then blindly 
searched for excess
power between 500 Hz and 1400 Hz using a scanning technique, as
presented in \citet{Boirin:2000jt}. We have also verified that no
significant excesses were detected between 1400 and 2048 Hz. This
justifies the use of the $1400-2048$~Hz range to estimate
accurately the Poisson noise level in each observation, which is 
indeed close to 2 in all segments of data.

The excess power is then fitted with a Lorentzian with three free
parameters; frequency, full width at half maximum, and amplitude
(equal to the integrated power of the Lorentzian). The Poisson
noise level is fitted separately above 1400 Hz. 68\% confidence errors on each
parameter are computed in a standard way, using a $\Delta\chi^2$ of $1$. 
Following
\citet{boutelier09mnras}, our threshold for QPOs is related to
the ratio (hereafter $R$) of the Lorentzian amplitude to its
$1\sigma$ error\footnote{The Lorentzian function used in the fit
is $\rm Lor(\nu) = A \times W / (2 \pi) / [ (\nu-\nu_0)^2 +
(W/2)^2 ]$, where A is the integrated power of the Lorentzian
from 0 to $\infty$, W its width and $\nu_0$ its centroid
frequency. The fitted function is linear in A, and therefore its
error can be computed using $\Delta \chi^2$ (e.g.
\citet{Press:1992yf}). The RMS amplitude is a derived quantity,
computed as RMS=$\rm \sqrt{A/S}$, where $S$ is the source count
rate \citep{van-der-Klis:1989kn}. In this paper, we have defined
$R=A/\delta A$, from which the error on the RMS is estimated as
$\rm \delta RMS = 1/2 \times RMS \times R^{-1}$ after neglecting
the term $\rm \delta S/S$ in the derivative of the RMS equation.}
($R$ was often quoted and used as a significance).  A conservative threshold
$R=3$ means that we consider only QPOs for which we can
measure the power of the Lorentzian with an accuracy of $3\sigma$
or more. Such a  threshold corresponds to a  $\sim 6\sigma$
excess power in the PDS for a single trial, equivalent to $\sim
4\sigma$ if we account for the number of trials of the scanning
procedure \citep{van-der-Klis:1989kn}. The integrated power of
the Lorentzian is then converted into a root mean square (RMS),
expressed as a fraction of the total source count rate. 

\begin{figure}[!h]
\epsscale{1.125}
\plottwo{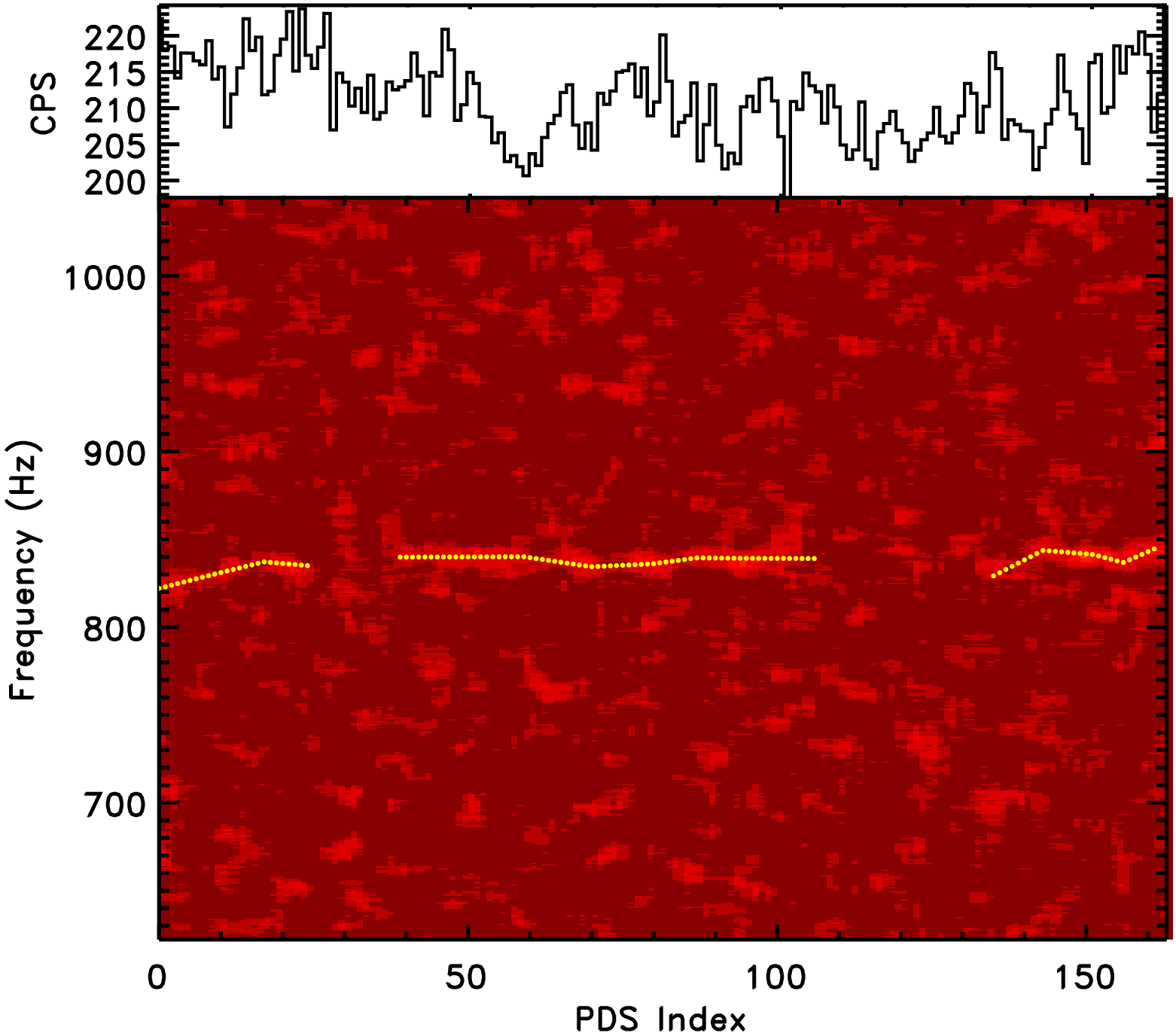}{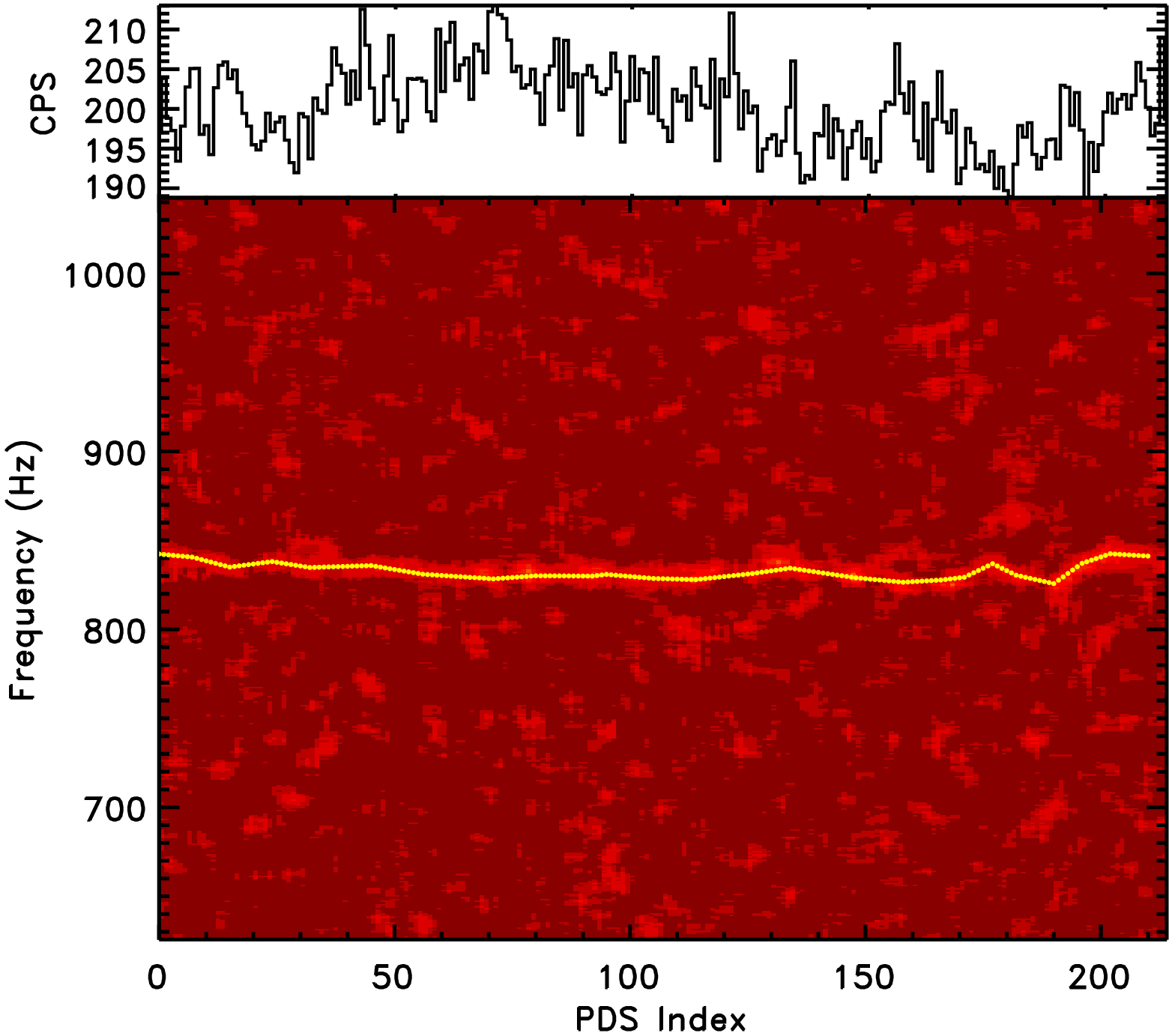}
\caption{The QPO frequency history recovered by our analysis for two
segments of observation of the ObsID 93703-01-02-05. The path
followed by the QPO is indicated with the yellow filled
circles. As can be seen on the left, there are gaps in the QPO detection. When the gap duration is less than 256 seconds, frequencies within the gap are estimated by interpolating between the two positive detections surrounding the gap. \label{fig1}}
\end{figure}

We detect a significant QPO in twelve segments of
observations, within the same ObsIDs as in \citet{sanna10mnras}.
The reduced number of active PCUs on the PCA, and the relative
faintness of the source in its atoll phase (count rate between
$\sim 40$ and $\sim 120$ counts/s/PCU), means that special care
must be taken to correct for the frequency drift of the lower kHz
QPO. We apply the very same technique as in
\citet{barret06mnras}. It is an iterative procedure, which
enables us to bound the QPO frequencies on shorter and shorter
integration time, with narrower and narrower frequency intervals.
As the integration time decreases, the significance threshold set
to the scanning technique \citep{Boirin:2000jt} is adjusted (e.g
the threshold is decreased from $4\sigma$ for an interval of 50
Hz over a few thousand seconds, to $3 \sigma$ for an interval of 10 Hz
over a few hundreds of seconds). The QPO path is then recovered
through a linear interpolation between the most significant
detections with the shortest integration time. An example of the application of the analysis to two 
segments of real observations is shown in Figure~\ref{fig1}.

Such a technique, which we have used extensively on real and simulated data, is able to correct precisely for the frequency drift of QPOs of the strength reported here. In those simulations, the QPO frequency evolution is modeled by a random walk of a given step (e.g. 0.2-0.5 Hz/second). We follow \cite{timmer95aa} and generate synthetic 1 second PDS, for which the underlying model consists of a constant (2) to account for the Poisson noise plus a lorentzian to model the QPO profile, with parameters (Q and RMS) appropriate for each frequency (estimated from the random walk). The simulated PDS are then combined on (16 second timescales) and scanned as the real data to recover the time evolution of the QPO frequency, and to determine Q and RMS, for comparison with the parameters injected in the simulations. The difference between the reconstructed and original Q and RMS parameters can then be evaluated. Those simulations have shown that for QPOs of comparable strength to the one of \xtej\ (mean RMS around 9\%), Q can be recovered with an accuracy better than 10\% (the accuracy on the RMS is better than a few \%).

\begin{figure}[!h]
\epsscale{0.6}
\plotone{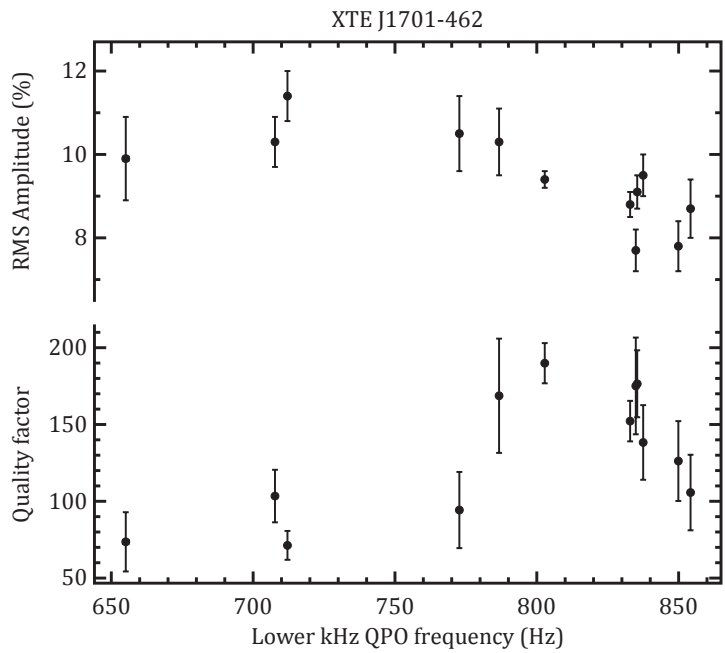}
\caption{RMS amplitude (top panel) and quality factor (bottom panel)
of the lower kHz QPO from \xtej~as averaged over the twelve segments
of observations. There is a trend for
the quality factor and the RMS amplitude to decrease after reaching
a maximum around 800 Hz, in a very similar way as observed in other
sources. \label{fig2}}
\end{figure}

Having reconstructed the QPO frequency evolution, one can compute
the mean QPO parameters over the observations, by
aligning all the 16 second PDS for which a QPO frequency was
estimated, directly from a significant detection or by interpolation between two significant detections, separated by less than 256 seconds, (see Figure \ref{fig1}). Increasing the latter value to say, 512 seconds, does not lead to any changes in the fitted QPO parameters.
The best
fit results are listed in Table \ref{tab1} and presented in
Figure \ref{fig2}. All the QPOs reported have a $R>5$,
and are therefore highly significant. As can be seen from
Figure \ref{fig2}, despite some scatter, there is already
evidence that above 800 Hz the quality factor and the RMS
amplitude of the QPO drops.

In addition to those highly significant QPOs, we also note that there
are hints for two additional single QPOs in the segments
93703-01-03-04 and 93703-01-05-06. No frequency drift corrections are
possible for those two QPOs. In the first segment, $\nu=905.0\pm4.4$
Hz, $Q=40.6\pm19.3$ for an $R$ ratio of 2.9, while in the second one
$\nu=548.3\pm33.2$, $Q=2.7\pm1.6$ for an $R$ factor of 2.3. Being
single and not very significant, it is difficult to draw any firm
conclusions. However, it cannot be excluded that they are lower kHz
QPOs, extending on both sides the frequency span of figure
\ref{fig2}, in which case one would expect them to have low Q
factors, as measured (with indeed the caveat that no drift correction
could be applied, implying that the measured values are only lower
limits on Q).

\begin{figure}[!h]
\epsscale{0.6}
\plotone{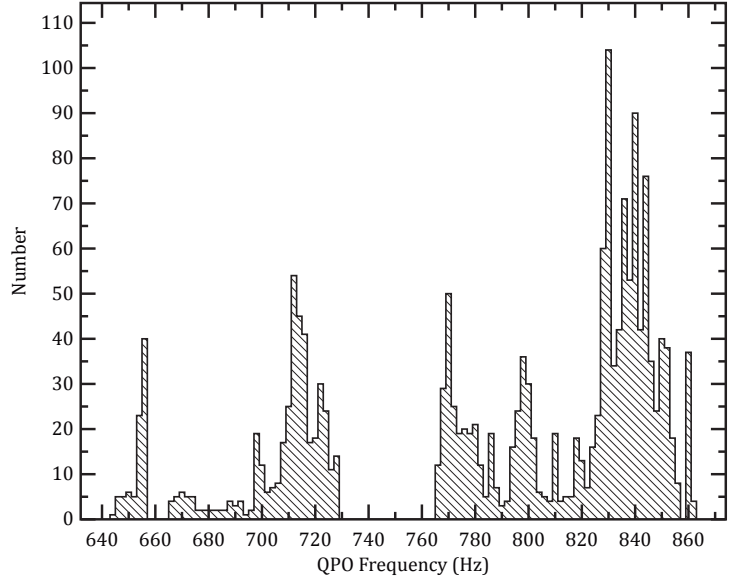}
\caption{Histogram of the frequencies (one frequency per 16 seconds)
detected in all 12 observations, showing that not all frequencies
have been sampled equally. The frequencies used for building the histogram are available in an online table. \label{fig3}}
\end{figure}

\begin{figure}[!h]
\epsscale{0.6}
\plotone{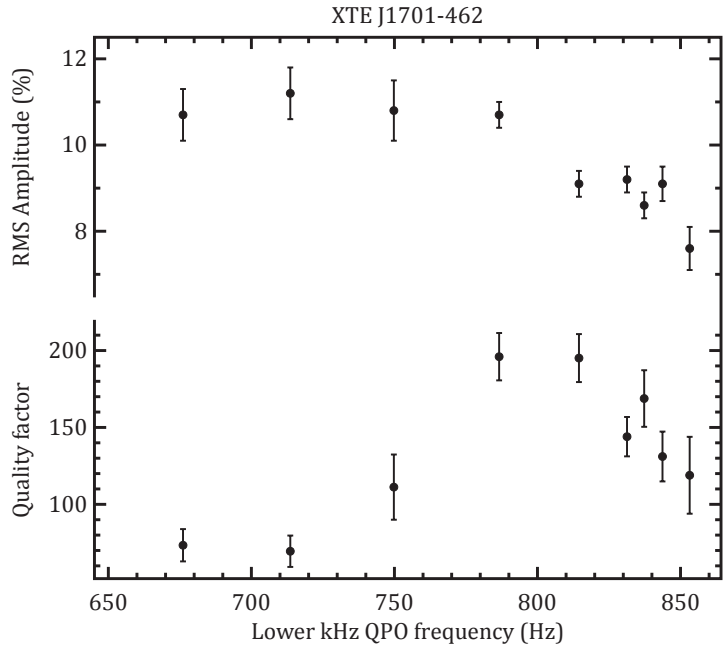}
\caption{Same as Figure 2, but the QPO parameters are averaged over
frequency intervals (200 consecutive frequencies were combined).
The drop in Q and RMS amplitudes are now clearer.\label{fig4}}
\end{figure}

The histogram of interpolated frequencies (measured over 16 seconds) 
is shown in Figure \ref{fig3}.
As can be seen, the frequency span of the lower QPO, although comparable
in breadth with other systems, has not been sampled equally (there is a
lack of observations around 750 Hz). Grouping the data over constant
frequency intervals is therefore not the optimum way to proceed. Instead
we have considered adjacent frequency intervals of varying widths, each including 200 QPO frequencies,
allowing a better sampling of the peak of the quality factor versus
frequency curve. The result is shown in Figure \ref{fig4}. As can be 
seen, the drops around
800 Hz of the QPO RMS amplitude and quality factor are now much clearer
and have significantly less scatter.
Thanks to our QPO tracking procedure, we can recover the QPO frequency
on a timescale of 16 seconds, whereas \citet{sanna10mnras} averaged as
many 16 second PDS as required to enable a significant detection. This
explains why on average we get larger Q factors than
\citet{sanna10mnras}, and hence a better description of its frequency
dependence.

\begin{figure}
\epsscale{.60}
\plotone{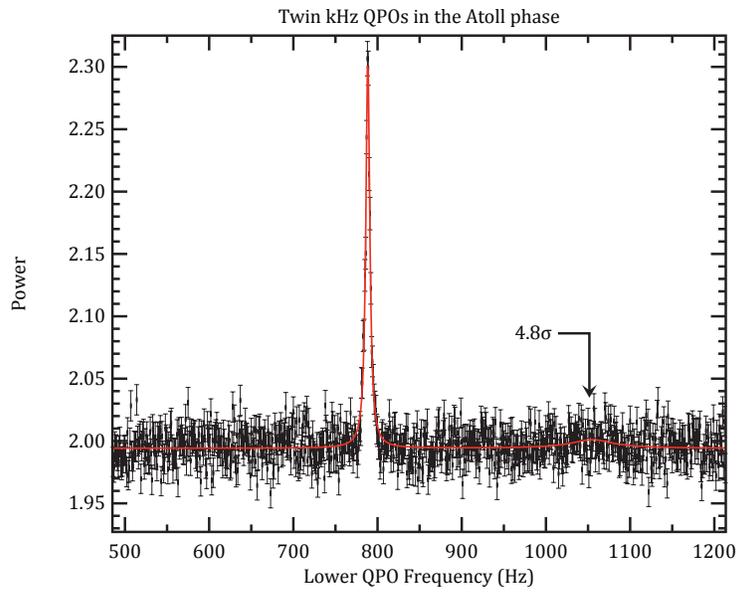}
\caption{The two kHz QPOs detected by shifting and adding all the 16
second PDS in which a QPO frequency could be estimated. The main
peak has a significance above 50$\sigma$, whereas the upper kHz QPO
has a single trial significance of 4.8 $\sigma$. Its quality factor
is $17.1\pm9.1$ and its RMS amplitude $4.4\pm0.8$, and its $R$ ratio
is 2.6\label{fig5}}
\end{figure}

Following \citet{sanna10mnras}, we have shifted all the 16 second
PDS to search for the upper kHz QPO. \citet{sanna10mnras}
reported a $3.1\sigma$ significance detection (single trial) with
a frequency separation of $258\pm 13$ Hz. Using our procedure, the 
significance of the upper kHz
QPO detected rises to 4.8 $\sigma$ (single trial significance),
for a frequency separation of $266.5\pm13.1$ Hz, consistent with
the value of \citet{sanna10mnras}.  At the same time, the significance
of the main peak increased from $30\sigma$ to more than
$50\sigma$ in our analysis (see Figure \ref{fig5}). In the two ObsIDs 
of the Z phase, in which the simultaneous twin kHz QPOs are the most 
significant ($R \ge 3$) (92405-01-40-04 and 92405-01-40-05), we have 
measured a frequency difference of $280.8\pm17.6$ and $276.1\pm18.8$~Hz 
respectively, i.e. consistent within errors with the frequency difference 
we have measured in the Atoll phase.

\section{Discussion}

Based on the observations reported here, our main findings can be summarized as follows:

\begin{itemize}
\item \xtej~behaves similarly to the other sources we have studied.
In the atoll state, its lower kHz QPOs are narrow. Correcting for the
frequency drift, we have been able to show that its quality factor
reaches a maximum of 200 around 800 Hz before dropping off sharply.
This effect is even visible within a limited sample of
observations, when the drift correction is applied. We have also
found a weaker upper kHz QPO with a significance greater than previously 
reported, using the shift-and-add technique. 

\item Following \cite{barret06mnras}, we estimate the frequency at 
the ISCO by adding $\sim 270$ Hz $\nu_{\rm lower,Q=0}$, which we infer 
to be around 900 Hz (albeit with large uncertainty due to the limited 
sample of the $Q-\nu$ curve in the available data set). This gives 
$\nu_{\rm ISCO}\approx 1170$~Hz. No upper QPO above that frequency 
should be detected in \xtej. This is consistent with the findings of 
\cite{sanna10mnras}, who reported upper QPOs with frequencies less 
than $\sim 930$ Hz. From $\nu_{\rm ISCO}$, we can estimate the gravitational
masses  of the neutron stars \citep{miller98apj}: \begin{equation}
M\approx 2.2\,M_\odot(1000~{\rm Hz}/\nu_{\rm ISCO})(1+0.75j)
\end{equation}
where $j\equiv cJ/(GM^2)\sim 0.1-0.2$ is the dimensionless angular
momentum of the star.  The inferred mass of the neutron star in \xtej\
would therefore be $\sim 2.0\,M_\odot$.

\item As shown by \cite{sanna10mnras}, when the source is in the Z phase,
the properties of its QPOs are consistent with those observed in other Z
sources: they are weak and broad (the maximum Q values are about 10 and
the RMS less than $\sim 4$\%). For the purpose of this paper, we have also
re-analyzed the segments of data in which \cite{sanna10mnras} reported a
QPO, and our results are globally consistent with theirs. Note however
that no frequency drift corrections can be applied when recovering the
quality factor, hence the measured values should be considered as lower
limits. In any case, our independent analysis confirms that no highly
coherent QPOs have been detected from \xtej~in the Z phase. This indeed
suggests that there is a parameter which alters the Q value, in addition
to the radius at which the QPOs originate.
\citet{sanna10mnras} conclude that these results weaken the hypothesis 
relating the drop in coherence observed in many sources, and in the Atoll
phase  of the one under investigation, to the presence of the ISCO,
because there are evidently other properties of the accretion flow that
can change the quality factor of QPOs.  We point out, though, that the
drop in coherence that we relate to the ISCO  is a very characteristic
phenomenon. In a given source, it happens always at  the same frequency,
at the higher end of the frequency range of kHz QPOs.  There is therefore
a significant difference between the drop in coherence we relate to the
ISCO and the change in coherence in the different states of this source.
In our opinion, this implies that different phenomena are producing the
two effects.

\end{itemize}

Let us now discuss on some possible interpretations. The change in coherence of the kHz QPOs in XTE~J1701-462 between the Atoll  and Z phase, has to be related to a global change of the accretion configuration.

In the framework of the toy model proposed by \citet{barret07mnras},
we argue that the difference is related to the scale height of the disk,
exactly the same process that we propose as responsible
for the general difference in coherence between QPOs in Atoll and Z sources.
Consistent with this idea, standard disk accretion theory suggests that at
rates approaching Eddington (which is expected in the Z state) the disk
half-thickness at its inner edge will be comparable to the orbital radius
there \citep{shakura73aa}, and that as a consequence the inward radial
drift speed (which scales as $(h/r)^2$, where $h$ is the disk
half-thickness) will be large as well. As discussed in
\citet{barret06mnras}, a large inward speed will necessarily decrease Q
regardless of other factors. It is therefore not surprising that high
luminosity sources have lower Q, as observed \citep{mendez06mnras}.  

We note that alternative explanations are under study through 
3D MHD simulations that show oscillations promisingly similar to the
kilohertz QPOs (for a recent discussion see \cite{bachetti10}).
In these simulations, the lower and upper kHz QPOs are emitted from
hotspots on the surface of the star, produced by streams of matter
originating respectively from the funnel flow around the magnetic pole and
from instabilities at the inner edge of the disk. The properties of
these QPOs in simulated systems with an enhanced variety  of physical
inputs (including different configurations of the disk) are  currently
under study.

\section{Conclusions}

The source XTE~J1701--462, which has been observed in both
the atoll and Z phases, gives us a unique opportunity to
explore the relation between frequency and quality factor.
This relation has been claimed for previous atoll sources
to provide evidence of the innermost stable circular orbit.
It was expected, however, that for higher-luminosity sources,
in which the disk becomes geometrically thick and the inward
radial speed is thus large, the quality factors would be
systematically lower than they are in the lower-luminosity
atoll sources, as observed \citep{mendez06mnras}.  XTE~J1701--462 
does show exactly this behavior.
In the atoll state its $Q$ vs. $\nu$ phenomenology is consistent
with that of other sources, but in the Z state the quality
factors are systematically smaller than in the atoll state.
The change is over the whole range of the QPO frequencies and is
unlike the abrupt drop observed in the atoll phase, which only 
happens at the higher end of the frequency range.  
As a consequence, we do not agree with \citet{sanna10mnras} that 
the different coherence in the atoll and Z phases of XTE J1701--462 weakens 
the ISCO hypothesis.

We can ask: what behavior would contradict
the ISCO picture? As discussed in other papers  
\citep[e.g.][]{barret07mnras}, there
are several possibilities including (1)~a lower or upper kHz QPO frequency
in another state that is significantly larger than the projected
maximum due to the ISCO, (2)~a peak in $Q$ followed by a sharp drop
at a frequency significantly less than seen in the atoll state,
or (3)~a peak in $Q$ followed by a sharp drop in any source at a
frequency that would imply an unreasonable ISCO frequency (e.g.,
below 800~Hz).  None of these have yet been seen.  Therefore, although
the importance of the implications demand further critical analysis,
the interpretation is still consistent with all existing data to date.

MCM acknowledges NSF grant AST0708424. The authors wish to thank an anonymous referee for helpful comments.
 

\begin{thebibliography}{21}
\expandafter\ifx\csname natexlab\endcsname\relax\def\natexlab#1{#1}\fi

\bibitem[{{Bachetti} {et~al.}(2010){Bachetti}, {Romanova}, {Kulkarni},
  {Burderi}, \& {di Salvo}}]{bachetti10}
{Bachetti}, M., {Romanova}, M.~M., {Kulkarni}, A., {Burderi}, L., \& {di
  Salvo}, T. 2010, \mnras, 403, 1193

\bibitem[{{Barret} {et~al.}(2005{\natexlab{a}}){Barret}, {Klu{\'z}niak},
  {Olive}, {Paltani}, \& {Skinner}}]{barret05amnras}
{Barret}, D., {Klu{\'z}niak}, W., {Olive}, J.~F., {Paltani}, S., \& {Skinner},
  G.~K. 2005{\natexlab{a}}, \mnras, 357, 1288

\bibitem[{{Barret} {et~al.}(2005{\natexlab{b}}){Barret}, {Olive}, \&
  {Miller}}]{barret05bmnras}
{Barret}, D., {Olive}, J., \& {Miller}, M.~C. 2005{\natexlab{b}}, \mnras, 361,
  855

\bibitem[{{Barret} {et~al.}(2006){Barret}, {Olive}, \&
  {Miller}}]{barret06mnras}
---. 2006, \mnras, 370, 1140

\bibitem[{{Barret} {et~al.}(2007){Barret}, {Olive}, \&
  {Miller}}]{barret07mnras}
---. 2007, \mnras, 376, 1139

\bibitem[{{Boirin} {et~al.}(2000){Boirin}, {Barret}, {Olive}, {Bloser}, \&
  {Grindlay}}]{Boirin:2000jt}
{Boirin}, L., {Barret}, D., {Olive}, J.~F., {Bloser}, P.~F., \& {Grindlay},
  J.~E. 2000, \aap, 361, 121

\bibitem[{{Boutelier} {et~al.}(2010){Boutelier}, {Barret}, {Lin}, \&
  {T{\"o}r{\"o}k}}]{boutelier10mnras}
{Boutelier}, M., {Barret}, D., {Lin}, Y., \& {T{\"o}r{\"o}k}, G. 2010, \mnras,
  401, 1290

\bibitem[{{Boutelier} {et~al.}(2009){Boutelier}, {Barret}, \&
  {Miller}}]{boutelier09mnras}
{Boutelier}, M., {Barret}, D., \& {Miller}, M.~C. 2009, \mnras, 399, 1901

\bibitem[{{Homan} {et~al.}(2007){Homan}, {van der Klis}, {Wijnands}, {Belloni},
  {Fender}, {Klein-Wolt}, {Casella}, {M{\'e}ndez}, {Gallo}, {Lewin}, \&
  {Gehrels}}]{homan07apj}
{Homan}, J., {et~al.} 2007, \apj, 656, 420

\bibitem[{{Homan} {et~al.}(2010){Homan}, {van der Klis}, {Fridriksson},
  {Remillard}, {Wijnands}, {M{\'e}ndez}, {Lin}, {Altamirano}, {Casella},
  {Belloni}, \& {Lewin}}]{homan10apj}
---. 2010, \apj, 719, 201

\bibitem[{{Leahy} {et~al.}(1983){Leahy}, {Darbro}, {Elsner}, {Weisskopf},
  {Kahn}, {Sutherland}, \& {Grindlay}}]{Leahy:1983mb}
{Leahy}, D.~A., {Darbro}, W., {Elsner}, R.~F., {Weisskopf}, M.~C., {Kahn}, S.,
  {Sutherland}, P.~G., \& {Grindlay}, J.~E. 1983, \apj, 266, 160

\bibitem[{{Lin} {et~al.}(2009{\natexlab{a}}){Lin}, {Altamirano}, {Homan},
  {Remillard}, {Wijnands}, \& {Belloni}}]{lin09apjb}
{Lin}, D., {Altamirano}, D., {Homan}, J., {Remillard}, R.~A., {Wijnands}, R.,
  \& {Belloni}, T. 2009{\natexlab{a}}, \apj, 699, 60

\bibitem[{{Lin} {et~al.}(2009{\natexlab{b}}){Lin}, {Remillard}, \&
  {Homan}}]{lin09apja}
{Lin}, D., {Remillard}, R.~A., \& {Homan}, J. 2009{\natexlab{b}}, \apj, 696,
  1257

\bibitem[{{M{\'e}ndez}(2006)}]{mendez06mnras}
{M{\'e}ndez}, M. 2006, \mnras, 371, 1925

\bibitem[{{Miller} {et~al.}(1998){Miller}, {Lamb}, \& {Psaltis}}]{miller98apj}
{Miller}, M.~C., {Lamb}, F.~K., \& {Psaltis}, D. 1998, \apj, 508, 791

\bibitem[{{Press} {et~al.}(1992){Press}, {Teukolsky}, {Vetterling}, \&
  {Flannery}}]{Press:1992yf}
{Press}, W.~H., {Teukolsky}, S.~A., {Vetterling}, W.~T., \& {Flannery}, B.~P.
  1992, {Numerical recipes in FORTRAN. The art of scientific computing}, ed.
  W.~H. {Press}, S.~A. {Teukolsky}, W.~T. {Vetterling}, \& B.~P. {Flannery}

\bibitem[{{Sanna} {et~al.}(2010){Sanna}, {Mendez}, {Altamirano}, {Homan},
  {Casella}, {Belloni}, {Lin}, {van der Klis}, \& {Wijnands}}]{sanna10mnras}
{Sanna}, A., {et~al.} 2010, ArXiv e-prints

\bibitem[{{Shakura} \& {Sunyaev}(1973)}]{shakura73aa}
{Shakura}, N.~I., \& {Sunyaev}, R.~A. 1973, \aap, 24, 337

\bibitem[{{Timmer} \& {Koenig}(1995)}]{timmer95aa}
{Timmer}, J., \& {Koenig}, M. 1995, \aap, 300, 707

\bibitem[{{van der Klis}(1989)}]{van-der-Klis:1989kn}
{van der Klis}, M. 1989, in Timing Neutron Stars, ed. H.~{{\"O}gelman} \&
  E.~P.~J. {van den Heuvel}, 27--+

\bibitem[{{van der Klis}(2006)}]{vdk06}
{van der Klis}, M. 2006, {Rapid X-ray Variability}, ed. {Lewin, W.~H.~G.~\& van
  der Klis, M.}, 39--112

\end{thebibliography}
\bibliographystyle{apj}

\newpage

\begin{deluxetable}{lllllllllllll}
\tabletypesize{\scriptsize}
\tablecaption{Lower kHz QPOs detected in the atoll phase of \xtej.
Parameters are averaged over the segments of observation, after
aligning all the PDS to a reference frequency. The table lists the
segment number, the ObsID name, the date, the duration of the
segment, the mean count rate normalized by the number of active
PCUs, the reference frequency of the QPO, its quality factor, its
RMS amplitude, and the accuracy $R$ ratio, as defined in the text.}
\tablewidth{0pt}
\tablehead{
\colhead{Num} &\colhead{ObsID} & \colhead{Date} & \colhead{Time} 
&\colhead{Duration} &\colhead{C} & \colhead{$\nu$} & \colhead{Q} 
& \colhead{RMS} & \colhead{$R$} 
}
\startdata
01&93703-01-02-04 & 2007/07/24 & 08--44--11 & 1968&116.9&$849.9\pm0.5$&$126.2\pm26.0$&$7.8\pm0.6$&7.0\\
02&93703-01-02-04 & 2007/07/24 & 10--15--27 & 1280&117.3&$854.2\pm0.7$&$105.7\pm24.6$&$8.7\pm0.7$&6.3\\
03&93703-01-02-05 & 2007/07/25 & 10--02--49 & 1920&105.8&$837.5\pm0.3$&$138.3\pm24.3$&$9.5\pm0.5$&8.8\\
04&93703-01-02-05 & 2007/07/25 & 11--23--27 & 3376&66.7&$832.9\pm0.2$&$152.2\pm13.2$&$8.8\pm0.3$&16.0\\
05&93703-01-02-08 & 2007/07/25 & 13--14--40 & 2304&104.9&$835.4\pm0.2$&$176.5\pm21.8$&$9.1\pm0.4$&11.0\\
06&93703-01-02-11 & 2007/07/25 & 05--13--35 & 2528&108.2&$834.9\pm0.3$&$175.1\pm31.5$&$7.7\pm0.5$&8.1\\
07&93703-01-02-11 & 2007/07/25 & 06--40--31 & 3392&68.1&$802.8\pm0.1$&$189.9\pm13.1$&$9.4\pm0.2$&20.4\\
08&93703-01-02-11 & 2007/07/25 & 08--14--23 & 208&71.0&$786.7\pm0.4$&$168.7\pm37.2$&$10.3\pm0.8$&6.3\\
09&93703-01-03-00 & 2007/07/29 & 03--19--30 & 3344&56.4&$772.7\pm0.8$&$94.3\pm24.8$&$10.5\pm0.9$&5.6\\
10&93703-01-03-00 & 2007/07/29 & 04--53--19 & 3408&39.2&$712.1\pm0.6$&$71.3\pm9.4$&$11.4\pm0.6$&9.7\\
11&93703-01-03-00 & 2007/07/29 & 06--27--27 & 2976&42.1&$707.7\pm0.4$&$103.4\pm17.1$&$10.3\pm0.6$&9.1\\
12&93703-01-03-02 & 2007/07/29 & 11--23--57 & 1360&39.7&$655.1\pm1.0$&$73.6\pm19.3$&$9.9\pm1.0$&5.2\\
\enddata
\label{tab1}
\end{deluxetable}

\end{document}